# ON VERIFICATION OF SOFTWARE COMPONENTS

Basem Y. Alkazemi

[1]Department of Computer Science, Umm Al-Qura University, Makkah, Saudi Arabia
bykazemi@uqu.edu.sa

## ABSTRACT

*Utilizing third party software components in the development of new systems became somewhat unfavourable approach among many organizations nowadays. This reluctance is primarily built due to the lack of support to verify the quality attributes of software components in order to avoid potential mismatches with system's requirements. This paper presents an approach to overcome this problem by providing a tool support to check component compatibility to a specification provided by developers. So, component's compatibility can be checked and developers can verify components that match their quality attributes prior of integrating them into their system.*

## 1. INTRODUCTION

Building software system according to the Component-based Software Development (CBSD) model become one of the widespread approaches in the software engineering field due to its significance to lower development cost and increase productivity. One of the common activities in CBSD approach is the composition of components to build new software system. This activity is solely dependent on investigating component's interfaces to identify their suitability for integration into a system. Component's interface exposes the abstract specifications of components and hides their implementation details. Thus, it must be carefully examined in order to avoid unforeseen problems that might be raised after the integration of software components [7].

The absence of systematic approaches for verifying and validating software systems is, in fact, one of the biggest challenges encountered in software development nowadays [4]. Essentially, detecting defects in the early stages can significantly reduce the overall fixing and maintenance cost. As a result, this project aims at establishing a software verification mechanism as a step forward to examine systems attributes in order to ensure that a system satisfies its exact needs and identify any unrelated business rules for investigation. It is commonly known that software developers are concerned with two main dimensions when building new systems, these are:

- Customer requirements

    - Functional
    - Non-functional

- System requirements

    - Syntactic
    - Semantic

Every dimension from the above incorporates a number of metrics that provide the baseline for measuring compatibility. So, a software system can be considered as a perfect-match solution for a problem domain if it satisfies all the characteristics of the identified dimensions. This work assumes that software components can be integrated in a system only if they provide the required behaviour and also conform to the system's structure at hand. This assumption is motivated by the fact that developers may find components that provide the right match to their required behaviour and use them into their system, but soon they discover that components failed to operate due to missing some of their required methods, for instance, or being configured according to different assumptions than the ones required by a developer.

We distinguish between two different types of components matching, behavioural and structural matching. This separation between behavioural and structural match can help to understand the characteristics of software component in more formalized manner. Achieving both types of matching is the ideal scenario where a component can be considered as an exact match to the requirements of the system at hand. We believe that satisfying the behavioural match is not enough to ensure that a component can be a fully operational one, hence this work concerns verifying some of the key characteristics of the structural match as a step towards achieving comprehensive verification model for software components. The structure that a component conforms to is named as its *physical properties*; it defines the values for a pre-defined set of characteristics that a component must satisfy in order to match the structure imposed by a system. These characteristics constitute part of component's interfaces. For example, one of the physical properties of a Java application is that it must contain a method called "`public static void main(args [])`". Based on our practical examinations of a number of software components, we identified that the characteristics defined by the physical properties of software components are distinct in nature and different from one structural type to another. As a result, we believed that it is useful to utilize the characteristics defined by components' physical properties to establish the basis for verifying component compatibility to system requirements.

The project will be discussed in seven sections, including this introductory one, followed by an explanation of the idea of the physical properties of components as defined in this project. The third section will look at the components compatibility scale, and the following two sections – four and five – will look at the verification and validation processes, and then discuss the approach in terms of design considerations of SpecJ. The related works is given in section six. Finally, in section seven, some conclusions will be offered, and suggestions for further work made.

## 2. The Physical Properties of Software Components

We refer to the physical properties of software components abstractly as the contract that allows components to work in a system. Both components and a system must agree upon a pre-defined contract in order to allow for a component to be integrated into a system. The contract is commonly represented by components interface. The characteristics defined by an interface capture the behavioural and structural aspects of software components. Based on the exhibited characteristics of a component's interface, a component can be identified and integrated. Component's interface can be represented directly in the code of the component (e.g. Java Interfaces) or by using additional files that describes the visible attributes and methods of the component.

Two types of interfaces are distinguished that we adopted the terminologies behavioural interface and structural interface. The behavioural interface exposes the set of services that a component can provide to their clients. Obviously, this interface is the key to identify if a component is of any interest to a developer as it is the first thing that can be examined. However, the behavioural interface is not so important when it comes to integration as the component will not be of any value if it cannot be integrated into a system; hence structural interface comes to the picture. The structural interface defines the physical properties of software components that if matched to a system's properties the component can work in that system. In fact, this interface is significant to examine aspects about components integration, as it is responsible for identifying whether components can physically fit into a system or not. For example, an Oracle 6i form will not fit directly into Microsoft SharePoint system due the lack of web-based support indicating a mismatch occurrence in the structural characteristics. Thus, the notion of structural interface is the main focus of this paper. Overlooking behavioural interface in this research does not mean that it is not of any importance, but our aim is to support software developers to refine their searching criteria with additional characteristics that are defined by structural interface.

Consider the `android.app.ApplicationContext.Application` class as a possible example of a system that a developer wants to add some functionality to it by incorporating new classes. The system provides an extensible environment that precisely defines how new classes can be added to the system, and also establishes the basis for defining the relationships between all the classes. The structural interface of the abstract class requires the following methods to be implemented by a generic Java class in order to fit in the android system:

```
protected void onCreate(Bundle savedInstanceState);
protected void onStart();
protected void onRestart();
protected void onResume();
protected void onPause();
protected void onStop();
protected void onDestroy();
```

These methods represent part of the structural interface that must be implemented by a class in order to fit into the android system. Based on examining various components types, a number of key physical properties are identified to characterize the structural interface of software components, they are:

- **Format** refers to the syntax that undergirds a component and which is used to write it. At the level of source code, for example, this refers to the programming language. Thus, if a given system has components written in Java, an additional component that is created in FORTRAN will not ultimately be suitable.
- **Entry point** refers to the code used in the first block of program that is used in initialization of a given component. Some components control the initialization of a process, or create a methodology for execution of a task, while others require specific files in order to operate. For example, a Java application designed to be "stand alone" requires "`public static void main ()`" for initialization, compared with plug-ins for android applications that start up using "`AndroidManifest.xml`". Components must use the same mechanism for initialization in order to ensure proper operation.
- **Fault Handler** is the term for a fault in a component within any stage in the running of a program, and the response of the software to that fault. If a system assumes that components will react with a particular "recovery action" when failures occur, a newly built component must have the same sort of reaction system.
- **Dependencies** are the components that use dependencies, which also need to align with the expectations of a given system. Java systems, for example, require composing components (called Java classes in this case) to run using a particular library termed "`java.io`" in order to correctly run any input/output functions. Android systems, on the other hand, use custom built library as their required components, for example "`android.intent.action.MAIN`", which controls for the execution of a function. Each component of a given system must thus use the same external dependencies.

- **Data I/O** comes into play once a given component of the software has been initialized and is ready to process received data, or send out data packages. How data is handled must align with the base system. The parameters for the reception of data, for example, include requirements for reading data from a file, and must fit with the existing input system. A single data-exchange model must be created, for example a push-model or a pull-model, since the two cannot be used within the same system's components.

- **Control flow** controls the exchange of component information, so that a given component is able to synch its execution with other components. This is done so that once the component has completed its task; control can be handed back over to the central system. In some systems, however, components work asynchronously, so how the execution of different components is carried out must be identified if they are to be integrated.

- **Design** is the invocation of a sequence of components that must be activated in a particular way in order to achieve optimal functionality. A compiler, for example, is made up of a reader component, which reads a file then takes the data to be stored in another temporary folder from where it will be buffered before it is processed. Then, an analyser component is activated that uses the data stored by the reader. If the analyser component is activated first in this sequence, there will be an error, and the system will fail to execute its desired action.

It is essential to identify each of the elements outlined above in order to understand and determine whether a newly designed component will physically fit and function within an already built system. A developer must first identify the characteristics and their values precisely, in order to ensure that the proper code is used. For this project, a prototype of specification language was developed, namely *SpecJ*, that describes components in relation to the identifiable syntactic constructions of their properties. This was carried out without regard to the behaviour of the component or the meaning of the construction. Take the Java Method, where the aspects of syntax that were identified had to be separated from the semantics that defined method (for example that result in a failure handling).

The main consideration of our designed language is to return a "true/false" answer with respect to matching a component against the characteristics defined by their corresponding physical properties. This consideration has the advantage of facilitating a tool to check automatically the availability of the characteristics in the structural interface of software components without so much of human intervention. The prototype of SpecJ has been developed mainly to formalize some of the characteristics identified by the structural interface, hence can serve as a verification mechanism that checks source code conformance to the required physical properties of a system. SpecJ is simply an XML-based document that defines the physical properties of a software component that can be utilized by a tool (e.g. ParseX) in order to examine the compatibility aspects of source code components. We are going to discuss the compatibility scale in the next section and identify where our intended specification language fit into that scale. Table 1 below lists the syntax of the SpecJ language.

Table 1. SpecJ Specification

| Tag | Description |
|---|---|
| `<SpecJ>` | Identify a document is SpecJ specification |
| `<SpecJ>\<name>` | Define the name of the type |
| `<Physical_Properties>` | Capture the properties of a component type |
| `<Physical_Properties>\<Block>` | Define memory address from source code |
| `<Block>\<name>` | Define memory name |
| `<Block>\<Data_Input>` | Define component input data stream |
| `<Block>\<Data_Output>` | Define component output data stream |
| `<Block>\<Failure>` | define exception handling mechanism |
| `<Block>\<File>` | define external file that architectural type use |
| `<Block>\<Storage>` | define temporary memory |
| `<Data_Input>\<sequence>` | identify the sequence of input data |
| `<Data_Output>\<sequence>` | identify the sequence of output data |
| `<sequence>\<type>` | define data type |
| `<Failure>\<type>` | define the type of exception handling |
| `<Dependencies>\<lib>` | define required resources |
| `<File>\<name>` | define the name of file |
| `<File>\<type>` | define the type of file |
| `<Storage>\<name>` | define memory address |
| `<Storage>\<type>` | define memory type |
| `<type>\<sub-type>` | define specialized file type of a generic one |

## 3. COMPONENTS COMPATIBILITY SCALE

Components compatibility ranges on a scale from "not compatible" to "Fully Compatable" as illustrated in figure 1.

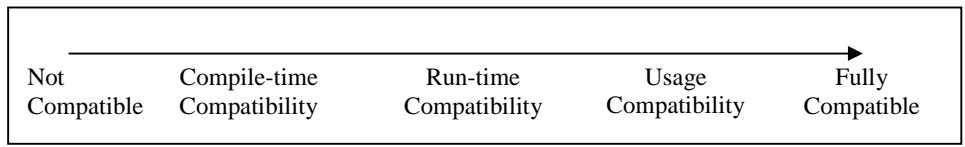

Figure 1: Compatibility scale

The first category on the scale, not compatible, is not of interest but we mentioned it here to indicate that source code component that does the required functionality might be found but does not fit in the system at hand. The compile-time compatibility category captures all the aspects that make certain source code components compile without problems. This is what most IDEs check for. Run-time compatibility includes aspects that make a system run without problems. For example, certain system requires its components to read from a file, or certain component must be a thread-safe. Usage compatibility category captures the aspects related to getting at component's functionality. For example, when a certain component might need to run, or what methods a component must have to be invoked at certain time, or where a component must be located so the system can find it, or registration is required before invoking component services. In fact, this category seems an important one as it tells the developer about how certain component can be utilized. The last category on the compatibility scale is the fully compatible; this is where components satisfy, in addition to the previous three categories, the behavioural requirements. This is not the concern of our SpecJ at this stage as it is related to the functionality of software component that we ignore in this work. Our proposed SpecJ captures the three categories in the middle of the scale (i.e. compile-time, run-time, and usage compatibility). In fact, the compile-time compatibility check is left to IDE's, so no detailed information are captured by SpecJ with this regard apart from specifying the primary language of certain software component and its dependencies, then IDE's can do the subsequent check. The run-time and the usage compatibility are the core criteria for measuring component compatibility in this work, therefore our SpecJ characteristics are influenced by these categories.

## 4. VERIFICATION PROCESS OF THE PHYSICAL PROPERTIES

We have developed tool called *ParseX* purely in Java using a common Java editor that check components against the physical properties definition represented in the SpecJ prototype. The approach followed by the ParseX tool for matching the physical properties to a provided component is based on utilizing the notion of compiler associated with the programming language to check the syntax of a component and also identify whether a component is missing any of its required sub-components. The tool works by automatically generating a "XClass" –an executable Java class– Java class from the provided specification document that describes the physical properties of the required components. The XClass class contains code to verify all of the features specified in the SpecJ document. The tool then compiles and links the generated Java class with the source code of the provided component. If no compile-time or link-time errors are raised, this indicates that the provided source code matches the physical properties that were used to generate the XClass, and the tool returns a positive result. If compile or link errors are raised, this reflects a mismatch and the tool returns a false match result. Figure 2 below illustrates the proposed verification process.

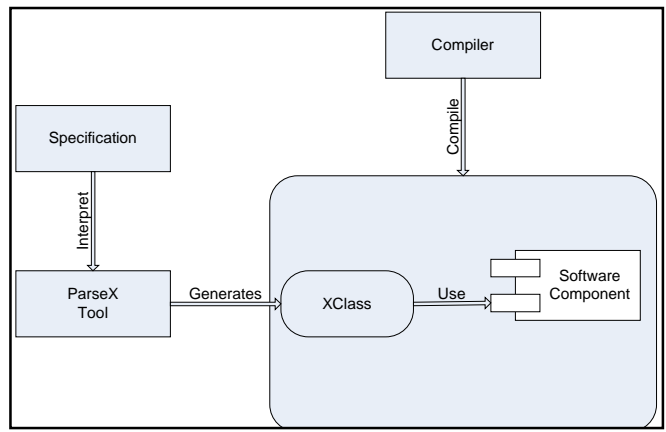

Figure 2: Verification Process

We have conducted some experiments to evaluate the applicability of SpecJ to capture the physical properties of some commonly known systems types. EJB has been selected as one possible type of systems for the evaluation of our approach. So, the experiment aims at verifying whether randomly selected open-source software components conform to the EJB physical properties, hence can be compatible with any EJB=based systems.

Text matching in open-source repositories was used to search for the string "EJB". A list of 312 results that contained the phrase "EJB" were returned as in 03/2012. We applied the ParseX tool against all the 312 components in order to identify if they match our SpecJ description of the EJB physical properties. SpecJ description for the EJB physical properties is given partially in Figure 3.

The ParseX tool identified that 301 of the 312 components matched the EJB SpecJ document. The remaining 11 components were flagged as not matching. To check the validity of the generated results, all the 312 components identified as EJB were examined on an EJB based system. It was found that all the 312 components ran successfully, including those components that the ParseX tool identified as being instances of the EJB type.

Figure 3: Simplified listing of the Physical Properties for EJB

Inspecting by hand the source code of the 11 components that returned negative result by the ParseX tool showed that all the components matched the basic properties defined by the EJB specification. After examining the possible reasons for the conflict in results obtained by the ParseX tool and by trying the components on the EJB system, the reason for the conflict was identified. The 11 components for which the negative results were returned by the ParseX tool seemed to be obtained lacking some of their required dependencies; assuming that they are going to be deployed in will provide these libraries for them. As a result, the compile-and-link process in the ParseX tool failed. This was the main reason that caused the ParseX to return negative results and not because these components were not conforming to the EJB specification. So, this result is considered a false negative result as the failure in the compilation was not caused due to missing any of the characteristics of the EJB specification but it was related to missing custom type dependencies that components require after their execution in an EJB based system. Despite these false negative results, the results obtained in this iteration are promising. However, this experiment uncovered some additional aspects that we may need to consider in the next version of SpecJ.

In summary, this experiment demonstrated that the notion of structural interface can be utilized for the purpose of verifying component compatibility at the structural level within the scope identified in this paper. The experiment showed that the defined characteristics of the EJB specification represented by SpecJ have been automatically checked against a number of real software components, hence the ParseX tool can be considered as a semi-formal verification mechanism.

## 5. SPECJ DESIGN CONSIDERATIONS

SpecJ captures several characteristics representing the physical properties that software components matched. The main design considerations addressed by SpecJ include the following:

- **Separation of concerns**: The expression "Concerns" represent the requirements that are fulfilled by a software system to accomplish certain task including performance, security, and functionality. The idea of separation of concern is not new in the field of software engineering. In object-oriented programming, the induction of classes enabled separating between different functional concerns within the component's source code. Whenever new key functionality is required, a class can be generated to capture a required behaviour. However, object-oriented languages still cannot separate between all kinds of concerns within component's source-code, especially those concerns that are related to non-functional requirements. For example, one may not able to tell which methods within a Java class are part of the implementation of a "bubble sort" functionality that is the main functionality of the class from the methods responsible of providing "logging" functionality within the same class. Aspect-oriented programming (AOP) established a mechanism to tackle the problem of separation of concerns that object-oriented approaches cannot address efficiently. AOP approach is fundamentally built on separating between different functional behaviours that are seen tangled in component's source code. AOP achieved the separation of concerns by introducing the notion of *point-cuts* that marks places within the source code to reflect potential concerns. For example, point-cuts could be placed at the beginning of a method's body to reflect starting method's execution and at the end to indicate finishing the execution. Whenever a point-cut is reached by the thread of execution special code is executed, this code is named as *advice* in AOP terminology. So, from the earlier example (i.e. bubble sort functionality and logging functionality), the "logging" source code that was mixed with "bubble sort" functionality source code in the case of object-oriented programming is substituted by advices in AOP. As a result, component's source code became more modular and maintainable as concerns are nicely separated. However, AOP is not concerned about capturing components structural interfaces, as a result, one may not able to tell which components source code is responsible of providing certain behaviour and which one is responsible for the physical matching of components into a system. For instance, AOP is not useful to tell what source code responsible of exchanging control and data among components and the source code responsible of performing a required computation within components. In fact, AOP primarily addresses developers needs in case of changing requirements after a system, or part of it, is been built but not their needs in case of incorporating components to build a system in the first place. Although AOP can separate between components concerns, it cannot help to establish how one component might be replaced from a system and how new component might be placed instead. In short, matching components together in a system is not a concern of AOP as it seemed that components compatibility aspects are left to be sorted out by programmers.
- **Data & control encapsulation**: Say an atomic component is found and wanted to be integrated in a system – an atomic component is a one that has no further sub-components. Components may describe methods to manipulate their data. They also may have states to preserve data either locally inside components or on external places (e.g. Database). Some components may further implement methods to enable reaching data within components in addition to other methods to pass control to them. SpecJ encapsulates data and control that constitute part of components' structural interface. For example, a component may implement methods called "`public void setX(int x)`" and "`public int getX()`" that must be used to manipulate some of its states data. Also, the component may implement a method called "`public void run()`" that must be invoked, by a system or other components, to start running the components, hence passing control to it. These methods are regarded as part of the structural interface description of the component. The methods indicate the allowable ways of data and control interaction by the component. So, a component can be seen as satisfying data manipulation requirements using, for instance, JavaBeans structural interface and also control can be passed to it using the corresponding life-cycle interface. Capturing these requirements precisely by the SpecJ helps developers to decide what modification is needed. So, they may either modify the components they found to achieve data manipulation and control passing in the way expected by the system, or modify their systems to satisfy requirements that components are expecting from a system. For example, if control must be passed to a component by invoking a method called "`public void run()`" then a system must adhere to that requirement and have the necessary invocation to that method, alternatively, if a system requires its components to implement a method called "`public void start()`" to control components execution then components must implement this method in order to fit.

- **Hide implementation details**: Hiding implementation details is an important aspect as far as CBSD is concerned, so developers can utilize components based on their defined interfaces without worrying so much about how their behaviour is achieve by implementation (e.g. what data structure is been used) as far as components fulfil the required functionality and fully compatible. Once a component that provides the required functionality is found, developers can utilize the attached SpecJ description to understand component's interface. If developers must know how the component they found accomplishes its behaviour then they need to inspect the internal source code of the component. So a system can interact with components through their structural interfaces, an advantage of this is to avoid any harm to the system in case components have been changed with other ones to meet new requirements (e.g. performance) as components interfaces are still unchanged. For example, say an implementation of "`writeToFile`" method is required to provide the functionality of writing text to a file in the disk. To plug the method as a component into a Java Application, the method must conform to Java standards and also be wrapped by Java class. So SpecJ will be simply represented, in this case, as the Java class, whereas the internal properties is the Java method "`writeToFile`" that is considered as an atomic sub-component.

- **Component modification**: SpecJ captures the structure of software components that by which they can work into a corresponding software system. Defining the structural types using SpecJ facilitates knowing what needs to be modified in order to fit components in different system types than the one they were originally taken from. The ability to perform such modification between system types can enhance components reusability as it bridges the gap between what developers need to what actually available for them. For example, if developers need to modify an "EJB component" to turn into an "Android System", and the SpecJ descriptions are known to them for both types then the modification can be applied accordingly. In fact, a tool might be utilized to perform the modification automatically. If component's functionality needs some modification, then some of the composing sub-components require replacing with others that meet new functionality. Modifying components functionality is similar to modifying system's functionality as a software component can be considered as a system by its own that is composed of a number of sub-components.

# 6. RELATED WORK

The ideal case in which a system can be accepted by a customer is if it provides the exact services and also performs as expected. Based on that, our distinction between behavioural and structural interfaces of software components is used to draw the context for our work. The verification and validation (V&V) must cover both types of system properties in order to certify a system for customer satisfaction. In practice, the V&V techniques yield a number of key attributes; including:

- typical development life-cycle stages
- model-driven engineering (MDE)
- reuse based Software Engineering
- requirement tractability
- product-lines engineering

We consider the above attributes as the framework for conducting the survey work in this section in order to ensure that the survey is comprehensive and bias free.

Cheon et al [1] introduced verifying and validating framework for investigating Programmable Logic Controller (PLC) characteristics in order to build safety-critical systems. The V&V tasks includes preparation of software plans and V&V procedures, a checklist review and fagan inspection [2], requirement tractability analysis, formal verification methods, and life-cycle based testing. The main objectives of their approach are to obtain the correct functionality, performance, reliability, and safety that satisfy the requirements of critical systems. They used the generated PLC to build a prototype of an engineered safety features-component control system (ESF-CCS) in the KINCS project. They claim that this framework matches the specification of critical systems. Cha et al [3] establish a software qualification model for verifying and validating PLC that is applied to Reactor Protection System prototype for Nuclear Power Plants (NPP) based systems. Their V&V model is characterized by defining inputs, tasks, and outputs for every phase of a software life-cycle. A new software analysis process based on the HAZard OPerability (HAZOP) methodology has been developed to improve safety attribute of critical software systems. Their experiments revealed the optimization of software V&V modelling and the usability of their approach to be extended for different software development contexts (e.g. railways, medicine). Gherbi et al [5] designed a domain model for Availability Management Framework (AMF) configurations, called MAGIC, that validates services compliance to applications requirements. The validated configuration overs the set of services to be managed by AMF in a system, types and relationships of services, and the deployment of services on clustered nodes. A tool that automatically generates AMF configuration for different service types to facilitate compatibility checking has been developed. Barbosa et al [6] established a methodology for evaluating COTS products by evaluating the impact of applying robustness te sts and fault injection tests. Their approach has been evaluated on RTEMS operating system for satellite applications. Their experimentations revealed the lack of wrapping interface that eliminate potential failure of a COTS component to spread across the other parts of a system. In addition, they have identified that the component lacks any failure detection mechanism. The outcome of their experiments indicated that some COTS products are not safe to use even though they might satisfy functional requirements. IEEE standard 1012-2004 [9] established a V&V framework in the context of the typical software development cycles. They consider V&V as a process model that encompasses a number of activities described by to three main dimensions; inputs, tasks, and outputs. The inputs are the set of prerequisites that must be satisfied for a process to launch. Outputs, nevertheless, are the outcome of executing a process. Tasks are the set of activities that must be accomplished internally to generate outputs based in the provided inputs. The key scope of this standard is to determine whether a product that resulted of a given activities conforms to the requirements for starting that activities in addition to its compliance to user needs and use -cases. For example, in requirement V&V, the task of generating requirement traceability table must have SRS as an input in order to deliver the desired traceability table as an output. The VERDE [10,11] framework that is still under development in a recent ongoing project owned by ITEA2 and targets V&V for real-time embedded systems (RTES). They identified that, in reality, V&V activities start only after implementation and integration is completed, hence fixing defects can be very costly and time consuming. As a result, the main objective of this project is to address the challenges of systematically verifying and validating non-functional characteristics of software systems at early stages of their development. NASA established a V&V approach in their product-line for examining the development of mission-critical software systems [12,13]. Their V&V framework assumes a number of key requirements to be satisfied in order to enable V&V:

- establish rigorous traceability
- the presence of detailed documentations
- source-code accessibility
- use small configuration units

Pesola [8] defines a framework for product V&V incorporated along with all the different stages of any software development life-cycle. The framework is composed of three main elements; phase, methods/techniques, and tools. Although these elements are somewhat abstract, it can be customized to suite the need of different business cases. Table 2 below illustrates a customization of Pesola's framework to suite the need of a typical software development life-cycle (e.g. waterfall).

Table 2: Pesola's framework customized to Waterfall life-cycle [8]

| Phase | Methods/techniques | Tools |
|---|---|---|
| Requirements definition | Rapid prototypes | - |
| Design | Model reviews and modelling guidelines | Enterprise Architect |
| Implementation | Defect detection in editor | Eclipse |
| | Static code analysis | FindBugs |
| Testing and integration | Model-based testing | Conformiq Qtronic |

It is obvious from the above surveyed work that V&V is significant to be incorporated into the software development process either as an embedded or a separate process that runs in parallel. It seems crucial that V&V activities are to be performed by independent parties those are not directly involved in the development process in order to identify defects more effectively, hence the notion of IV&V was coined. Moreover, V&V framework must consider all the development stages of a software development process whether the development adhered to standard models or was done in an ad-hoc manner. All the work discussed above seem to focus on V&V activities that are conducted for software development organizations, but customers approval are not addressed in any of the presented work. It might be obvious that customers may not be aware of the technical details of the non-functional characteristics of the system they are in charge to approve, however, any failure that might encounter in the delivered system will be referenced to their decision. As a result, a V&V decision support mechanism must be established to support customer's approval of software systems. Moreover, the V&V process for software components is limited to safety and some high level quality attributes and not the physical properties as described by our approach, hence our approach can be a step forward in that direction.

# 7. CONCLUSION AND FUTURE WORK

This paper has presented a new approach for verifying software component. Our experimental work revealed that the proposed approach is efficient to verify the key architectural characteristics of software components. The paper has demonstrated how architectural interface can be utilized to verify software components prior to integration. In fact, this work uncovers a significant research direction that need to be considered in depth for successful component integration at the source code level. The current approach is limited to small to medium sized software components written in Java. As a result, our planned future work is to apply the notion of architectural interface to verify a wider range of physical properties including components written in different programming languages. Moreover, we are going to investigate the potential of verifying components that are partially matching the characteristics defined by architectural interface as the current approach is limited only to exact match.

We are targeting organizations in the region of Saudi Arabia to benefit from this study as the current state of software development in Saudi Arabia is still in its infancy, with very limited attempts to develop small to medium size software in house. Most of the software systems are either bought or outsourced to external vendors. The situation might get worse when it comes to building a complex system (e.g. ERP) that is composed of several components and every component is provided by a different vendor as numerous integration problems might be encountered. In addition, vendors may find gaps to sneak away from their duties for fixing problems that rose at the boundaries (e.g. the integration points). Currently, validation are done only by customer during the acceptance test stage, but other aspects of validation (e.g. performance, security) or even verification are not done properly. As a result, there is a crucial need to establish and adapt IV&V metrics within the region of Saudi Arabia in order to overcome the lack of customer satisfaction and enhance the infrastructure that paves the way for new generation of electronic systems (e.g. e-health, e-learning, e-government). Therefore, one of our next planned work is to apply an advanced version of SpecJ to some organizations within the region as a tool to verify their underlying systems composite.

## Biography

Basem Y. Alkazemi is an assistant professor at Umm Al-Qura University (UQU) in Saudi Arabia under the school of computer science & Engineering. He obtained his PhD in 2009 from Newcastle University in U.K. His PhD topic was concerned with addressing the complexity of re-using open-source software components. Basem is currently holding the position of vice dean of IT deanship for e-government at UQU. One of his main duties is to establish a framework that leads to the integration of all the university software systems in a unified model. He is a member in the IEEE, SIGSOFT-ACM, and SEI societies. His main research interests include software architectural patterns, software product lines, Aspect-oriented SE, SOA, and CBSE.